# Intelligent Computer Numerical Control Unit for Machine Tools

**Joze BALIC**
University of Maribor, Faculty of Mechanical Engineering
Laboratory for Intelligent Manufacturing
Post: SI-2000,    City: Maribor Country:  Slovenia
E-mail: joze.balic@uni-mb.si  http://www.uni-mb.si

**Abstract**

The paper describes a new CNC control unit for machining centres with learning ability and automatic intelligent generating of NC programs on the bases of a neural network, which is built-in into a CNC unit as special device.  The device performs intelligent and completely automatically the NC part programs only on the bases of 2D, 2,5D or 3D computer model of prismatic part. Intervention of the operator is not needed. The neural network for milling, drilling, reaming, threading and operations alike has learned to generate NC programs in the learning module, which is a part of intelligent CAD/CAM system.

**Keywords:** Intelligent CNC & CAD/CAM, Neural Networks, Machining Centres

## 1. Description of the problem

Figure 1 shows the conventional CNC unit, connected with the machining centre [1]. The input of data respectively NC program of the part is usually done through DNC (Direct Numerical Control) connection**,** with punched tape through the tape reader**,** or manual, through interface for manual input. The manual input is used only for corrections of NC program and for changing technological parameters.

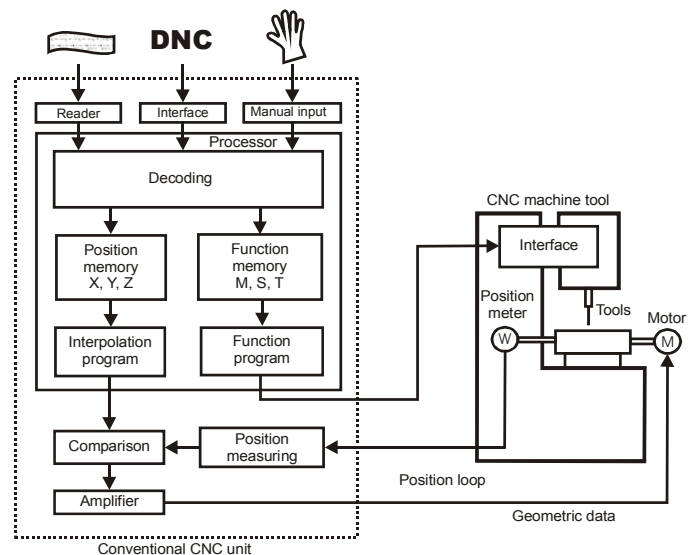

Figure 1:  The conventional CNC unit for machine tool

For each individual part the new NC part program must be set up, as the control does not remember already performed operations and can not automatically change program parameters (except some technological parameters, as are cutting conditions, corrections of lengths of tools, displacement of offset and zero points).

Also the use of modern CAD/CAM systems does not solve this problem. These systems enable that a new NC part program can be performed faster and more reliable. Some systems enable also saving of certain processing strategies, but the intervention of skilled NC programmer is still necessary. The programs performed on such way are not possible to use directly in CNC control of the machine tool, however they must be adapted in so-called post-processor phase. The task of this phase is that generally valid file of the tool path (CLDATA) is modified for each individual machine tool respectively CNC control. Each NC part program and each change must go over this post-processor phase.

## 2. State of the art and existing solutions

There are only some solutions known from the literature, which describe the broader field of research presenting in this paper.

Literature [2] describes the NC control unit with integrated function of learning. The NC control unit performs the teaching NC part program, which is compared with the inserted NC part program and performs then the resulting NC part program. On this way the operator of the machine tool can choose the "teaching mode" and changes the actual NC part program according to the suggestion from teaching program. The solution requires the intervention of the skilled worker during the NC programming.

Literature [3] describes a device for generating the tool path on NC machine tools and adequate NC control. The device at first recognizes the geometric feature characteristics of CAD model of the part and on the base of preserved processing procedures (machining cycles, sub-programs) chooses the most suitable tool path. The device can choose only that machining procedures, which have been previously defined as typical processing for particularly sub-programs.

Literature [4] describes the dialog orientated program system for programming of CNC machine tools. Hereby it is possible to choose interactively different control programs and procedures, which are then automatically composed into a NC part program. The intervention of the operator, programmer is necessary.

Literature [5] describes a learning method of a purpose made device. For this reason a special man-machine interface, which enables a dialog with the user and learning, is built-in into the control unit of the machine.

Literature [6] describes a device and a method for generating of NC programs. A special device saves the data about parts, belonging coordinates, characteristically junctions and time of assembly for single electronic components. The solution enables shortening of the time for the composition of NC programs and reduction of mistakes at preparing of programs.

In the paper [7] an adaptive controller with optimisation function of the milling process is described. It used neural network to adjust the learning procedure and for on-line modelling of the milling process. The efficiency of NN based controller is higher than of the conventional CNC controller.

In the paper [8] the new concept of CNC control unit is proposed. It consists from feature based NC unit and a basic control unit. The feature-based unit is used as exchange for geometrical data between basic NC unit and CAD/CAM system. It can be connected in Internet and used in virtual manufacturing.

Common for all described solutions is, that manual intervention of skilled operator, programmer is still necessary for preparing of NC part program for CNC machine tools. The systems can't prepare NC programs for parts, which are not saved in the databases and can't choose and use the machining strategies automatically.

## 3. Intelligent computer numerical control unit
### 3.1 Description
The new control unit for CNC machining centres - milling, drilling and operations like drilling is shown on Figure 2. It has the capability of learning and intelligent, automated generating of NC part programs on the bases of neural network, which is built-in into a

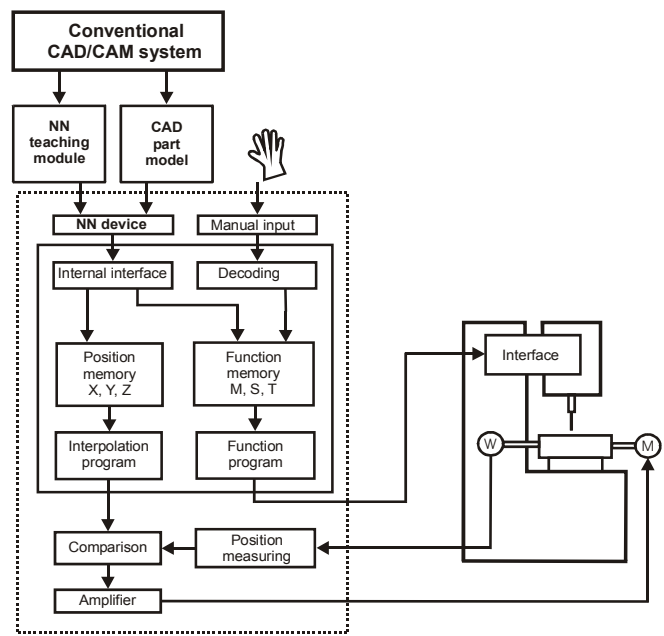

Figure 2: Neural Network Based CNC Machine Tool Control Unit

The neural network for milling, drilling, reaming, threading and operations like this have learned to generate NC part programs in system for learning, which is not a constituent part of a new CNC control unit, and works independent.

The new control unit then completely automatically, without any intervention of the operator, on the bases of a CAD model of a part, generate NC part programs for different products, which are designed for the production on CNC machining centres and for which previously NC part programs have not been done.

CNC control unit consists of a modified microcomputer, which includes a module for decoding and a module for manual input of commands, which are mostly of technological nature (feed rate, revolution speed, switch on/off of cooling liquid etc.). Computer includes also the internal interface, which has the function of data transmission from the NN device into the position memory.

### 3.2 Description of the NN device
NN device (Figure 3) consist of:
- A module for recognizing of geometric and technological features from CAD model of the part,

- Computer CAD model of the part, which contains geometric and technological features,
- NN device for milling, which has been learned for NC programming of milling, drilling and machining operations like drilling and
- NC control program for the part, which has been sent to control device, as a CAD model of the part.

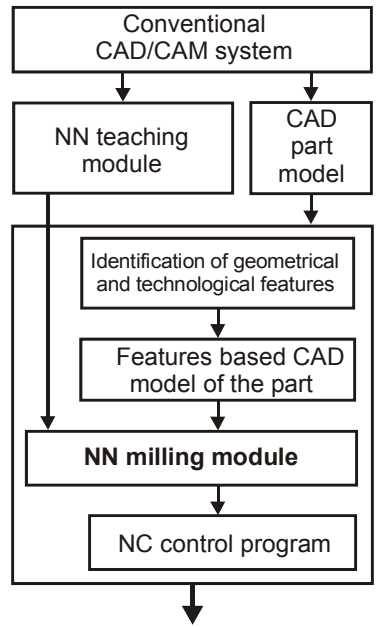

Figure 3: Schematic lay-out of NN device

In the phase of learning the NN device is connected to the module for learning of the neural network, which takes the data from the usual, commercial available CAD/CAM system, which is used for programming of NC/CNC machine tools. NN device automatically generates NC programs for CAD models of each individual product, which it becomes from the commercial CAD/CAM programming system. CNC control unit after this invention can receive the data for teaching of NN device from a special module for teaching, which is not the constituent part of the control unit. The task of the module for teaching is to teach the neural network, which is built-in into NN device, the principles and the technology of NC programming of all machining operations on CNC machining centres, mostly milling, drilling and operations alike.

In principle for this purpose different systems of neural network and different, also commercial developed software, can be used [9,10,11]. If we want to meet special criteria at machining (costs, time, quality of cutting, tool life, high speed cutting etc.) special developed neural network must be used.

## 4. Mode of operation
### 4.1 Intelligent automatic mode
In the mode of intelligent, completely automated processing the CNC control unit receives the data about CAD model of the product from the usual, commercially available CAD/CAM system for programming of CNC machines [13,14,15]. The model is then transmitted to NN device, which first of all recognizes geometric and technological features of CAD model of the part [16,17]. After recognizing the geometric and technological features are classified. On this way the new CAD model of the part is built again, which is now based on this characteristic features. Such model is then transmitted to NN device for milling, which on the base of learned intelligent procedures determines the most suitable machining operations and cutting parameters (cutting speed, feed-rate and the depth of cutting), with respect to chosen conditions (machining time, surface quality, machining costs).

The exit from the NN device for milling is NC part program for the processed part, which includes all geometric data about the mode (linear G01 or circular G02/G03 interpolation, etc.) and coordinates of cutting tool path (for instance milling cutter), technological data (revolution speed Sxxxx, feed-rate Fxx, depth of cutting) and auxiliary data (coordinates of reference, zero and starting points, direction of rotation of the main spindle M02/M03, change of cutting tools M06, etc.).

The data are then transmitted to internal interface, which split the data in NC program on tool path data (coordinates of movement in axis X, Y, Z and/or rotation A, B, C around coordinate axis X, Y, Z and on data about functions (M, S, T).

The geometrical data are obtained from the NC part program for each individual part and are treated in the position control circuit.

**4.2 Neural network learning mode**
In the learning mode the learned NC programming system, which is based on the principle of neural network, is put into the NN device, through the system for learning. The method of teaching the NN device is performed in a special NN device for teaching. The operation principle of NN device is shown more detailed on Figure 4.

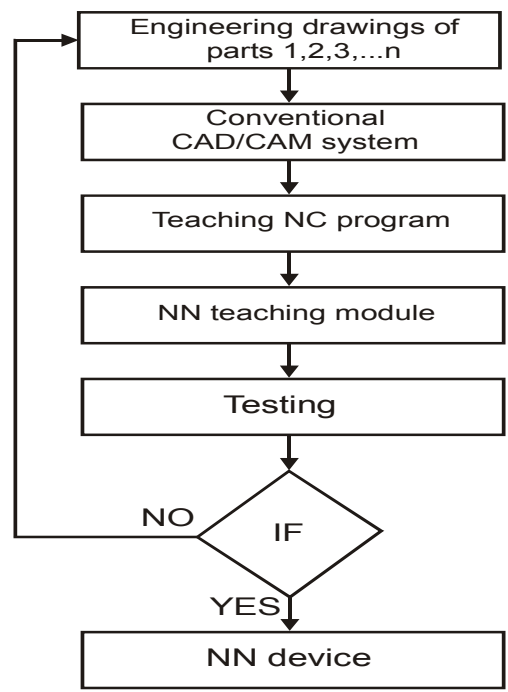

Figure 4: Procedure of learning and generation of neural network

In the learning phase, the NN device is connected to the teaching module designed for instructing the neural network (NN). The teaching module takes the data from conventional, commercially available CAD/CAM system for programming the NC/CNC machine tools. By means of a conventional CAD/CAM system, the teaching NC programs are prepared for different parts, defined in engineering drawings module, and are sent to the teaching module. In the decision module, subsequent to the testing module, the decision is taken on the success of teaching. In case that the decision is NO the repetition of the teaching process takes place. If on the other hand the NN device has learned enough the generated neural network is sent to the NN device.

The neural network built-in in the NN device consists of three layers: the input layer, the hidden layer and the output layer [12]. On the input layer, the X-Y-Z sets of coordinate points appear, representing the coordinate point values obtained from the modified CAD model for individual machining operations types. The coordinate point values are determined according to special procedure. Through the intermediate hidden layer the input coordinates are transposed into output layer in a form of a set of coordinate points $X_i$, $Y_i$, $Z_i$, representing the position values of the tool path for individual machining operations.

### 4.3 Machining operations

The control unit can learn how to generate the NC part programs for the following machining procedures: milling (rough face milling, rough contour milling, final milling after the contour and in Z-plane, final contour milling, final contour 3D milling, milling on Z-plane, final contour milling on equidistant and milling of pockets), drilling (normal drilling, deep drilling and centering), reaming, sinking and threading.

## 5. Industrial applicability

The NN device can be built-in into each CNC control unit of the milling machine as shown in Figure 1. The standard parallel transmission of data is used. If it is not possible to programming the internal interface the NN device must be connected to existing DNC interface, which each CNC control possess. The NN device for teaching is connected to NN device with standard serial interface. The CAD model of the part is sends to NN device through standard communication interface.

For teaching of NN device through the module for teaching different commercial available CAD/CAM programming systems can be used as are Unigraphics Solution, I-Deas, Catia, HyperMill etc. [13,14,15].

## 6. Conclusion

The new control unit solves the problem of automatic and intelligent generating of NC programs for CNC machining centres. It describes the control device of a CNC machining centre for face milling (rough), contour milling (rough), final milling after the contour and in Z-plain, final contour 3D milling, contour final milling, milling on Z-plain, final contour milling on equidistant, milling of pockets, normal drilling, deep drilling, centering, reaming, sinking and threading with the capability of teaching and automatic intelligent generating of NC part programs on the bases of neural network, which is built-in into a special NN device. The neural network for milling, drilling and operations alike can learn to generate NC part programs in module for learning**.** The device then completely automatically, without the intervention of the operator, only on the bases of 2D, 2,5D or 3D computer model of prismatic part**,** which has not been previously programmed, generate NC part programs. In solving the machining problems, the NN can serve as ideal tool, for helping NC programmer making the right decision, or at least serve as an orientation tool. Time saving could be also achieved, because not so many post-machining operations would be necessary.